# Training artificial neural networks for precision orientation and strain mapping using 4D electron diffraction datasets

Renliang Yuan[a,b], Jiong Zhang[c], Lingfeng He[d], Jian-Min Zuo[a,b]

[a] Department of Materials Science and Engineering, University of Illinois at Urbana-Champaign, Urbana, IL 61801, USA

[b] Materials Research Laboratory, University of Illinois at Urbana-Champaign, Urbana, IL 61801, USA

[c] Intel Corporation, Corporate Quality Network, Hillsboro, OR 97124, USA

[d] Idaho National Laboratory, Idaho Falls, ID 83415, USA

**Abstract**

Techniques for training artificial neural networks (ANNs) and convolutional neural networks (CNNs) using simulated dynamical electron diffraction patterns are described. The premise is based on the following facts. First, given a suitable crystal structure model and scattering potential, electron diffraction patterns can be simulated accurately using dynamical diffraction theory. Secondly, using simulated diffraction patterns as input, ANNs can be trained for the determination of crystal structural properties, such as crystal orientation and local strain. Further, by applying the trained ANNs to four-dimensional diffraction datasets (4D-DD) collected using the scanning electron nanodiffraction (SEND) or 4D scanning transmission electron microscopy (4D-STEM) techniques, the crystal structural properties can be mapped at high spatial resolution. Here, we demonstrate the ANN-enabled possibilities for the analysis of crystal orientation and strain at high precision and benchmark the performance of ANNs and CNNs by comparing with previous methods. A factor of thirty improvement in angular resolution at 0.009° (0.16 mrad) for orientation mapping, sensitivity at 0.04% or less for strain mapping, and improvements in computational performance are demonstrated.

**Highlights**

• A new approach to map crystal orientation and strain in TEM samples is described using trained artificial neural networks.

• It uses simulated dynamical electron diffraction intensity as the training dataset.

• The method is applied to the analysis of four-dimensional electron diffraction datasets.

• High angular resolution at 0.009˚ for precision orientation mapping and strain sensitivity at 0.04% are demonstrated.

## 1. Introduction

Real crystals contain various defects. Colin Humphreys, whose career we are celebrating here together with those of John Spence and Knut Urban, famously stated "Crystals are like people: it is the defects in them which tend to make them interesting" [1]. Examples of technological importance are many, such as, dopant atoms are used to control the electronic properties of semiconductors, dislocations underly crystal plasticity, and vacancy defects give rise to ionic conductivity. Real crystals are traditionally imaged using the so-called diffraction contrast in transmission electron microscopy (TEM) [2]. With the development of high-resolution electron microscopy (HREM), to which John Spence and Knut Urban have dedicated a large part of their research careers, the interruption of crystal lattice by dislocations or stacking faults can be observed directly [3, 4]. Impurity atoms and atomic disorder, under favorable conditions such as very thin crystals, can also be imaged [5-7]. Together, diffraction contrast imaging and HREM have contributed to much of our experimental knowledge about defects. However, unlike X-ray diffraction, quantitative analysis of real crystals has always been a challenge for TEM. Basic crystal information, such as the thickness and orientation of a crystalline sample, is only obtained under favorable conditions such as single crystal diffraction using convergent beam electron diffraction [8] and ultra-thin samples using quantitative HREM [9]. Here, we explore the possibilities offered by recent progress of artificial neural networks (ANNs) for precision diffraction analysis through the design and benchmark of ANNs for orientation and strain mapping applications.

ANNs are statistical learning algorithms that are modeled loosely after how human brain recognizes patterns. An ANN is typically organized in hierarchical layers, patterns are presented to the network via the input layer, which communicates to one or more 'hidden layers'. The hidden layers are then linked to an output layer where the answer is presented. Each layer in an ANN is made up of multiple processing elements, also called nodes or neurons, which are interconnected with nodes in the neighboring layers. The connection between two different nodes is assigned a numerical value, called weight. The output from one layer is used as input to the next layer. By systematically tuning the weights through an optimization process, the network can accurately approximate an arbitrary function. Once an ANN is structured for a targeted application, it must be trained using machine learning. In the so-called supervised learning, a set of labeled data containing both input and output (label) are presented to an ANN. The network is trained by comparing the output of the ANN against the desired output and updated by back propagation, in which the system adjusts the weights. This process is then repeated to optimize the weights. One of the major breakthroughs in the field of machine learning is the invention of convolutional neural networks (CNNs), which are ANNs containing multiple convolutional layers in addition to fully connected layers as in traditional ANNs [10]. CNNs are capable of deep learning to progressively extract higher level features from the raw input, for the recognition of complex patterns.

The abilities of machine learning, including deep learning, using ANNs present exciting opportunities for the analysis automation and augmentation of electron microscopy data [11-13]. In computer vision, deep learning has been used to solve difficult problems, such as classification, segmentation and detection [14]. In the field of electron microscopy, deep learning has been applied to, or proposed for, crystal symmetry determination in electron backscatter diffraction (EBSD) [15], the crystallographic analysis of electron image and diffraction data [12], resolution enhancement in scanning electron microscopy (SEM) images [16], defect analysis using simulated electron images [17], single atom detection [11, 13], and matching of experimental and simulated position averaged convergent-beam electron diffraction (PACBED) pattern for the determination of crystal thickness and tilt [18]. On the other hand, unsupervised machine learning methods, for example non-negative matrix factorization and clustering, have also been demonstrated their usefulness in electron diffraction through automatic learning of microstructural features contained in 4D data without the need of model training[19, 20].

This paper focuses on machine learning techniques for precision electron diffraction pattern analysis. The targeted electron diffraction patterns are electron nanodiffraction patterns obtained using a small focused beam. The motivation is to train ANNs for regression analysis based on the information of individual diffraction disks, their intensities and positions, instead of the whole diffraction patterns as done in the case of position averaged convergent beam electron diffraction (PACBED) [18]. As the size of input data reduces, the structure of ANNs can be simplified and the number of parameters to be optimized can be reduced, which provides significant benefits regarding the computational requirements and the reduced complexity of ANNs. For the training, we used the simulated dynamic electron nanodiffraction patterns. Experimental diffraction patterns are recorded from a sample region using the SEND technique [21], which is also known as 4D-STEM when it is performed in a STEM [22]. These techniques collect a 4D diffraction dataset (4D-DD) with two reciprocal space coordinates $(k_x, k_y)$ and two real space coordinates $(x, y)$. Automated analysis of 4D-DD is a central part of electron diffraction imaging. Here, we introduce the design, testing and benchmarking of ANNs for two applications of SEND. The first is precision crystal orientation mapping using a trained ANN, where we determine small changes in crystal orientation within a crystalline grain. In the second example, a CNN is trained to measure the position of electron diffraction disks to map crystal strain fields. For testing and benchmarking, the results obtained with trained neural networks are compared with the established correlation analysis technique [23, 24] for orientation mapping and the circular Hough transform (CHT) method [25] for strain mapping. To demonstrate the application potentials of trained neural networks, we apply our methods to image grain subdivision inthe nuclear fuel material of $UO_2$ after irradiation and a fin field effect transistor (FinFET) device. Using these cases, we demonstrate the steps required to take the full advantage of neural networks for electron diffraction.

## 2. Methods
### 2.1. Scanning electron nanodiffraction

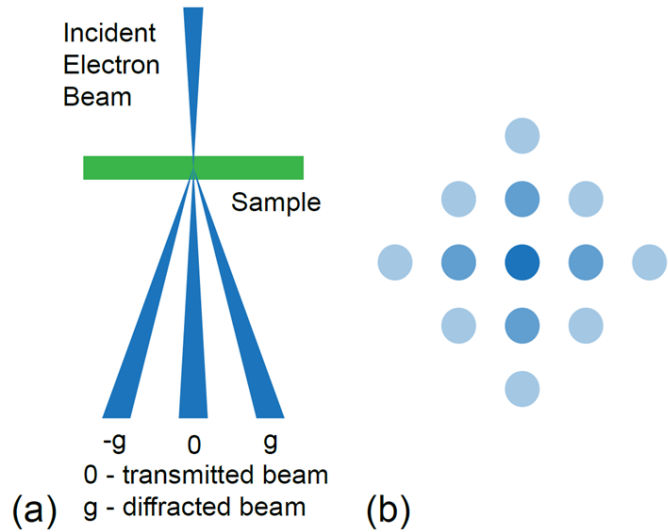

**Fig. 1.** Electron nanodiffraction using a small focused beam as illustrated by the schematic diagrams of (a) diffraction geometry and (b) diffraction pattern.

The types of electron diffraction patterns we analyze here are electron nanodiffraction patterns collected using the scanning electron nanodiffraction (SEND) technique [21, 26]. Here, we provide a brief summary on the key points of this technique.

Electron nanodiffraction patterns are recorded using a focused electron probe at a thin sample (Fig. 1). The beam convergence angle is in a few mrads, which yields small diffraction disks in the recorded patterns. The electron probe size, in the absence of lens aberrations, is diffraction limited according to $d_{FWHM} = 0.52 \frac{\lambda}{\theta}$ for the full-width half maximum (FWHM) probe size. At $\theta$ = 1 mrad and $\lambda$ = 1.9 pm for the 300 kV electrons, and $d_{FWHM}$ = 1 nm. Sub-nm probes can be obtained by increasing $\theta$ to a few mrad, in which case larger diffraction disks are recorded as in convergent beam electron diffraction (CBED).

The information recorded in a diffraction pattern can be quantified and categorized based on 1) diffraction geometry, in the form of Bragg-diffracted beams whose position and arrangements are related to the crystal unit cell structure and crystal orientation, and 2) diffraction intensity, which is related to the crystal structure, electron beam energy, sample thickness and sample orientation. Nanodiffraction patterns from a thin crystal show very little feature within each diffracted disk. This leads to the possibility of using the average, or integrated, disk intensity for structural analysis [27]. As the convergence angle increases, more complex diffraction patterns are recorded in thicker crystals due to the larger angular range and deviations from the Bragg diffraction condition [28]. In the applications here, we use dynamical diffraction for intensity prediction to improve the measurement accuracy.

The spatial resolution of electron nanodiffraction is ultimately limited by the electron probe size $d_{FWHM}$ and the cone diameter under the column approximation. The diameter of the cone $d_{AB}$ in a thin sample is defined approximately as

$$d_{AB} = 2\alpha t, \quad (1)$$

where $\alpha$ is the convergence semi-angle and $t$ is the sample thickness. The radius of the first Fresnel zone $\rho_1$ used to represent the diffraction column is calculated using

$$\rho_1 = \sqrt{\lambda t}, \quad (2)$$

where $\lambda$ is the electron beam wavelength. At 300kV, $\lambda$ = 1.9 pm. If the convergence semi-angle is 1 mrad and sample thickness is 80 nm, $d_{AB}$ is 0.2 nm and $\rho_1$ is 0.4 nm.

### 2.2. TEM samples

To demonstrate the applicability of our methodology as well as to test the method's reliability, we selected following samples. A thin single crystal sample of GaSb with the bending contour contrast, an irradiated polycrystalline $UO_2$ sample with grain subdivision and associated small angle boundaries, and a FinFET transistor device where SiGe is introduced to generate strain fields [25]. The $UO_2$ sample was prepared from a spent light water fuel from Belgium Reactor 3 (BR3) with an average burn-up of 4.5 at% [29]. The GaSb was mechanically thinned and then polished using ion beam milling. Both the $UO_2$ and FinFET samples were prepared by the focused ion beam (FIB) methods.

### 2.3. Experimental diffraction data collection

We acquired electron nanodiffraction patterns using a Themis Z S/TEM (Thermo Scientific, Waltham, USA), installed at University of Illinois. The microscope was operated in the μProbe STEM mode with the acceleration voltage of 300 kV. The electron probe focused on the sample had a semi-convergence angle of 1.2 mrad, and the probe size of 1.0 nm in FWHM. For precision crystal orientation measurement of GaSb, camera length was set at 185 mm, where about 40 diffraction peaks adjacent to the center beam in the [1$\bar{1}$0] zone axis were included in the recorded patterns. For strain mapping in the FinFET device, camera length was set at 360 mm so that only 8 diffraction peaks adjacent to the center beam were included. Diffraction patterns were recorded using a CMOS camera (Ceta, Thermo Scientific) at the resolution of 1024×1024 pixels and 0.1 s exposure time per diffraction pattern. The 4D-DD acquisition was automated by a control software provided by Thermo Scientific. During the scan, a STEM image, containing the sample region, was also acquired using the high-angle annular dark-field (HAADF) detector after each row scan for sample drift correction. The 4D-DD on irradiated $UO_2$ was collected using a Talos F200X S/TEM (Thermo Scientific, Waltham, USA), installed at Idaho National Laboratory, in μProbe STEM mode at 200 kV. The probe semi-convergence angle was 0.9 mrad and the probe size was 1.9 nm in FWHM. The camera length of 98 mm was used to obtain a large number of diffraction peaks adjacent to the center beam along the [112] zone axis direction. The detector setting in this case was same as the one used for GaSb.

### 2.4. Simulation of electron diffraction patterns

To build the diffraction pattern libraries for ANN training, electron nanodiffraction patterns were simulated using the Bloch wave method. The Bloch software was used for this purpose [8]. Three sets of diffraction library were built. The first one is for crystal orientation determination, where diffraction patterns of single crystal GaSb were simulated in a tilt series up to 0.5 degree tilt from the [1$\bar{1}$0] zone axis at a step size of 0.02 degree. The crystal tilt is defined by two tilt angles ($tilt_x, tilt_y$), which refer to the incident beam angle to a selected zone axis in degrees. They are used to calculate the tangential component of the incident wave vector

$$\vec{k_t} = \frac{tilt_x}{\lambda} \cdot \frac{\pi}{180} \vec{g} + \frac{tilt_y}{\lambda} \cdot \frac{\pi}{180} \vec{h}, \quad (3)$$

where $\lambda$ is the electron beam wavelength, $\vec{g}$ and $\vec{h}$ are two orthogonal unit vectors in the reciprocal space. We have assumed the tilt angles are small as in our experimental cases. The tilt series calculations were repeated for the crystal thickness from 65 to 85 nm at a step size of 5 nm. Both the tilt and thickness ranges were chosen to cover the estimated tilt and thickness variations in the GaSb sample. In total, 13,005 diffraction patterns were simulated.

Another simulation library was built for the orientation determination of UO$_2$. Diffraction patterns of single crystal UO$_2$ were simulated in a tilt series up to 2 degrees tilt from the [112] zone axis at a step size of 0.1 degree. The tilt series was repeated for the crystal thickness from 50 to 90 nm at a step size of 10 nm with a total of 8,405 diffraction patterns.

The third library was calculated for the d-spacing determination in a FinFET device. Diffraction patterns of single crystal Si with the thickness of 10, 15, and 20 nm were simulated in a tilt series with up to 0.6 degree tilt along the [110] and [001] directions from the [1$\bar{1}$0] zone axis at a step size of 0.3 degree. The thickness was selected based on the estimated thickness of the crystalline SiGe in the FinFET device. The camera length was set so that the radius of the diffraction disks is 45 pixels, same as the SEND experiment. A total of 43 sub-images of 121-by-121 pixels with each containing a single diffraction disk were cropped out from each simulated diffraction pattern. Among all the simulated diffraction disks, a library of 10,000 disk images were randomly selected to train the convolutional neural network. Random disk shift was applied to each disk image. We also added random noise to the disk images using the additive white Gaussian noise (AWGN) method with a varying signal-to-noise ratio (SNR) of 15~30 dB to mimic the diffuse scattering and detector noise in the experimental patterns. Examples are shown in Fig. 2.

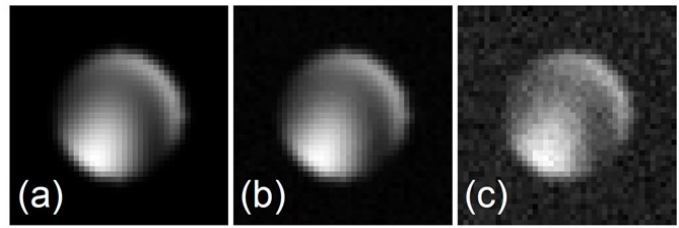

**Fig. 2.** Gaussian noise added to mimic experimental diffraction patterns. Simulated diffraction disks with (a) no noise, and with noise level of (b) 30 dB and (c) 15 dB signal-to-noise ratio (SNR).

## 2.5. Training artificial neural networks for precision crystal orientation mapping

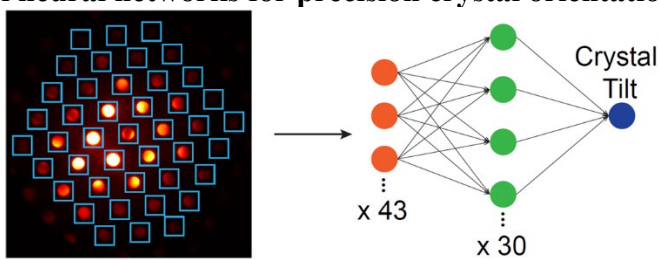

**Fig. 3.** The design of an ANN for precision crystal orientation determination based on integrated diffraction intensities. In the diffraction pattern example here, 43 blue boxes centered on the diffraction disks are used to calculate the integrated diffraction intensity as input to the neural network.

An ANN for regression was built for GaSb precision orientation mapping using 4D-DD. The simulated diffraction patterns are 1024×1024 pixels in size. For the training data, we integrated the diffraction intensities of 43 reflections as marked with blue boxes in Fig. 3. The blue boxes sit on a 2D grid, which were adjusted manually to make sure all diffraction disks reside close to the centers of corresponding boxes. In this way, the input data is greatly reduced from a 1024×1024 image to 43 integrated intensities. To mimic the variations in experimental diffraction intensities, we also added the Gaussian noise to the integrated intensities with a varying SNR of 15~30 dB. For the ANN, we used a simple three-layer model, which contains an input layer of 43 integrated intensities, 30 neurons in the hidden layer using the Sigmoid function as activation function, as shown in Fig. 3. The output layer determines the crystal tilt angle along X∥[001] direction in the [1$\bar{1}$0] zone axis pattern without nonlinear activation. Tilt angle along Y∥[110] is determined by another ANN with the same structure as tilt X. The ANNs were trained with the Levenberg–Marquardt backpropagation algorithm [30] using the 13,005 simulated diffraction patterns of GaSb with different orientations near [1$\bar{1}$0] and thicknesses, as described in Section 2.4. The training dataset contains a range of different thicknesses to help the ANNs to minimize errors in orientation determination due to sample thickness variations. Both input (disk intensity) and output (tilt angle) were normalized to 0~1 for better training performance.

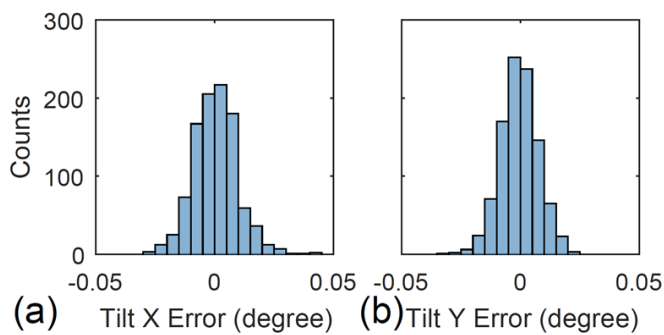

**Fig. 4.** Theoretical accuracy of orientation determination using ANN tested by simulated diffraction patterns. Histograms of measurement error in (a) tilt X and (b) tilt Y.

To evaluate the accuracy of orientation determination using the trained ANNs, we simulated another 1,000 diffraction patterns as test data with random crystal tilts up to 0.5 degree from the [1$\bar{1}$0] axis and random thicknesses from 65 to 85 nm. The same level of noise was applied to the test data and the training data. The difference between the simulated test patterns and the training library for the ANN is that the tilt angles and thicknesses in the test patterns are no longer on grids with the steps of 0.02 degree in tilt and 5 nm in thickness. The trained ANNs are applied to the test data to measure the orientation. The deviation in tilt angles from their true values are defined as error, the error distribution is plotted in Fig. 4. The accuracy of this method can be estimated by the standard deviations of the error, which are 0.0092 degree (0.16 mrad) in tilt X and 0.0078 degree (0.14 mrad) in tilt Y, respectively.

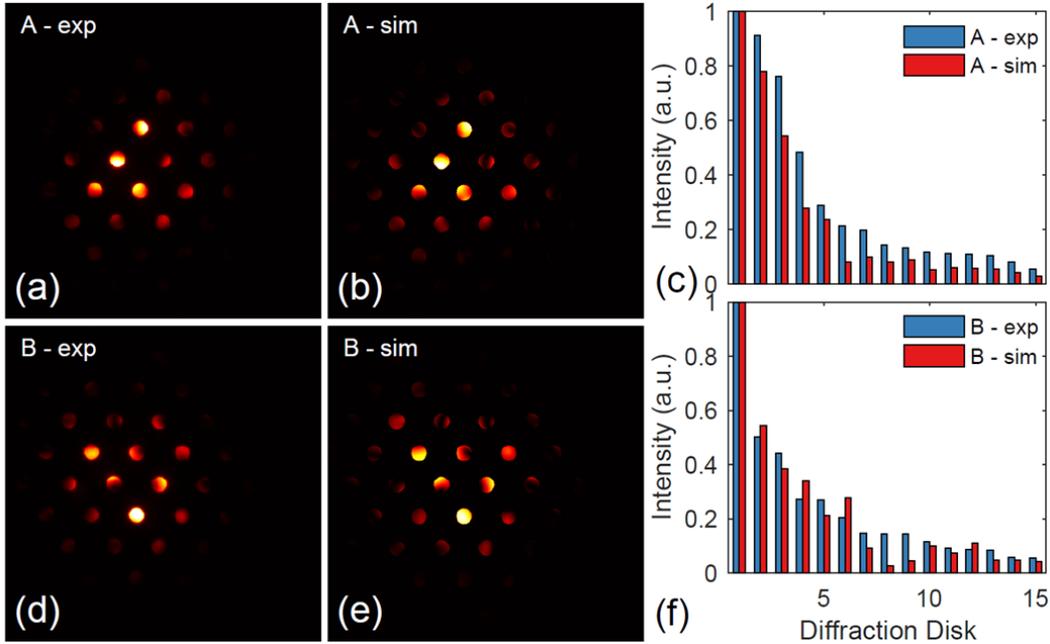

**Fig. 5.** Comparison between experimental GaSb diffraction patterns and simulated diffraction patterns with crystal orientation determined by the ANNs. (a)(d) Experimental patterns, (b)(e) best matching simulated patterns as identified by the ANNs, and (c)(f) integrated intensity of 15 brightest disks sorted according to the intensity value for experimental and best matching patterns of A and B, respectively.

To test the performance of ANNs on real experimental data, we chose two diffraction patterns, A and B from the GaSb sample, which are separated from each other in the 4D-DD. The crystal orientation was first determined by the trained ANNs. For pattern A (Fig. 5a), the tilt angles from [1$\bar{1}$0] axis are -0.2010° and -0.0537° along X and Y, respectively. The tilt angles of pattern B (Fig. 5d) are determined to be -0.2724 ° and 0.0947° along X and Y, respectively. Then, we simulated diffraction patterns with these determined tilt angles, and compared them with the experimental patterns with the crystal thickness of 80 nm (which gives the best match). The results are displayed in Fig. 5. The good matching in both Fig. 5a, b and Fig. 5d, e indicates the high accuracy of the trained ANNs, even though the intensity variations in the diffracted disks were not taken into account when the integrated disk intensities were used for the ANN training. Quantitative comparison of the integrated intensities from the experimental and simulation patterns is made in Fig. 5c and f, where the intensities of 15 strongest diffraction disks, normalized by the highest intensity, are plotted. The difference between experiment and simulation is quantified by the R-factor,

$$R = \frac{\sum |I_{exp} - I_{sim}|}{\sum I_{exp}}, \quad (4)$$

which is $R = 26\%$ for Fig. 5c and 19% for Fig. 5f, respectively. This shows that the above method is robust even though the experimental integrated intensities do not match the simulation exactly.

The ANNs for precision orientation determination of the irradiate $UO_2$ sample were designed using the similar approach as the ones designed for GaSb, except that in this case 35 integrated diffraction intensities were selected from patterns near the [112] zone axis. As the range of orientations we want to cover is much larger than the GaSb case, the number of hidden layers was increased from one to three, which improved the accuracy of the regression from 0.45 degree in *tilt*$_x$ and 0.43 degree in *tilt*$_y$ to 0.29 degree in *tilt*$_x$ and 0.24 degree in *tilt*$_y$. These estimations were based on 1,000 simulated diffraction patterns of $UO_2$ of random orientation within 2 degrees tilt from the [112] zone axis and random thickness between 50 and 90 nm.

## 2.6. Training convolutional neural networks for strain mapping

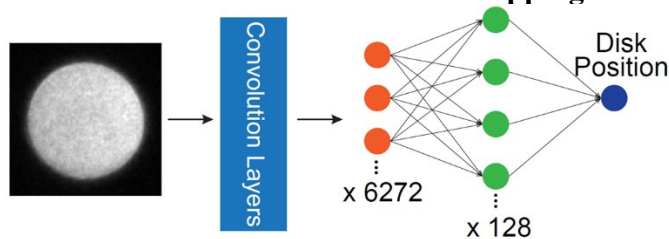

**Fig. 6.** Convolutional neural networks for diffraction disk position determination. The input is a 121×121-pixel image containing a single diffraction disk.

The principle of strain mapping is based on a determination of local d-spacing using electron nanodiffraction. The d-spacing is measured from the disk positions in the recorded diffraction patterns and by applying Bragg's law [25]. Critical to the whole process is to measure the positions of diffraction disks. This task is nontrivial due to the uneven intensity distribution that is typically observed in recorded diffraction disks, which is caused by dynamical diffraction.

The CNN model shown in Fig. 6 was built in TensorFlow. The input layer takes in the disk images of 121×121 pixels in this case, followed by two sets of 3×3×8 convolution layer with ReLU activation and 2×2 max-pooling layer combination, which help to extract high-level features from the input images and reduce the size. Then a fully connected layer of 128 neurons with Sigmoid activation is used to calculate the output of disk position X from the center of the image. The simulation library of 10,000 randomly shifted diffraction disks was used as training data. Adam [31] was used as the optimizer and the mean squared error was used as loss function.

Although the disk position we try to measure is two-dimensional, displacements in horizontal and vertical directions are independent to each other. In order to measure the disk position in both directions, we can simply apply the ANN to each disk image twice, once to the original image, the other to the image transposed. Considering the mirror symmetry associated with determination of disk position along one direction, the disk image is also flipped upside down for the second measurement of the horizontal displacement and left to right for the second measurement of the vertical displacement. Two measurements using the mirrored images are averaged to reduce the error.

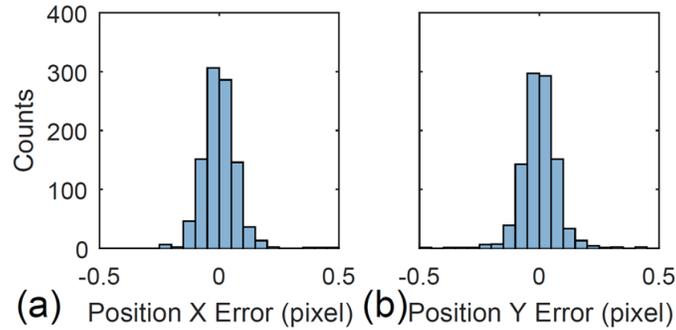

**Fig. 7.** Theoretical accuracy of disk position determination using CNN tested by simulated diffraction disks. Histograms of errors in the determined disk positions along (a) horizontal and (b) vertical directions.

The accuracy of the trained CNN for disk position determination was estimated by applying the trained CNN to a new batch of 1,000 simulated diffraction disks randomly selected from the simulation library. The same level of random displacements and noise were added to the new batch as the training data. The error distribution is displayed in Fig. 7. The standard deviations of the error are calculated to be 0.070 and 0.093 in pixel in horizontal and vertical directions, respectively.

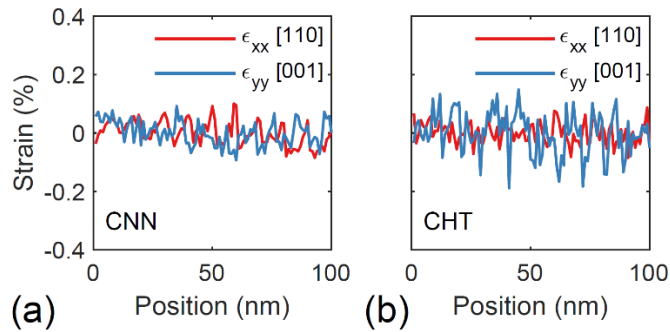

**Fig. 8.** Strain profiles from a SEND scan on a flat Si sample. Strain is measured by detecting diffraction disk positions using (a) the trained CNN, and (b) the CHT method.

Once the disk positions are determined, we follow the methods described in Ref. [25] to obtain strain. The precision of strain measurement using the trained CNN as a disk detection method was evaluated using a calibration SEND scan on a flat Si sample. The strain profiles calculated by CNN method are plotted in Fig. 8 along with those calculated by the CHT method [25] using the same dataset. The standard deviation of the strain in the calibration scan is taken as the upper limit of strain measurement precision, which is 0.041% along the [110] direction and 0.042% along the [001] direction for CNN results, and 0.036% along [110] and 0.076% along [001] for CHT results, respectively. The improvement of precision in strain along [001] shows the advantage of CNN method that it takes into account of the intensity of ±(400) diffraction disks, which because of its large scattering angle fluctuates from one pattern to another. The comparable precision along [110] is attributed to the robustness of both methods and the residual strain in the sample.

## 3. Applications
### 3.1. Precision orientation mapping of a GaSb thin sample

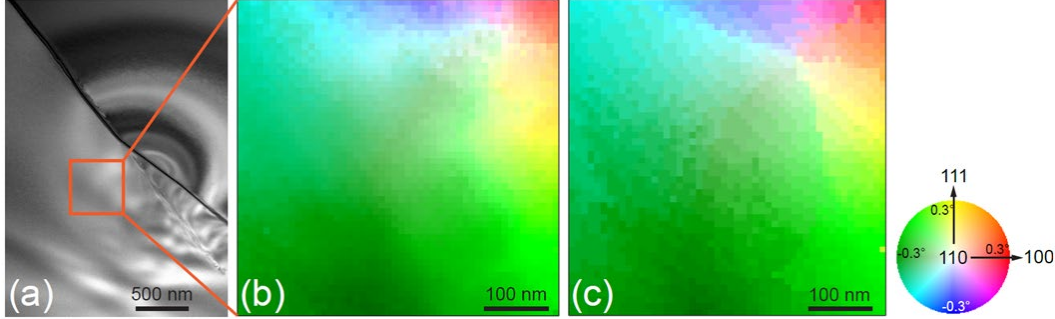

**Fig. 9.** Precision orientation mapping of a thin GaSb sample. (a) Annular dark-field image (ADF) showing diffraction contrast near a crack in the sample. Orange box marks the area of the 4D-DD was collected. (b) Orientation maps are calculated from the 4D-DD using (b) the ANN method and (c) the pattern matching method.

To demonstrate the sensitivity of precision orientation mapping using trained ANNs, we first applied the method to a hand polished GaSb thin sample in which bending is observed near a defect. Fig. 9a shows an image acquired using the same optics as SEND by an annular dark-field (ADF) detector with collection angles of 31~185 mrad. Dramatic diffraction contrast is seen in the ADF image, which can be attributed to crystal bending (local orientation variation) and thickness variation. A 4D-DD is acquired over a region of 500×500 nm$^2$, with a step size of 10 nm. Each electron nanodiffraction pattern in the dataset is reduced to 43 integrated disk intensities as the input for the ANNs trained for crystal orientation determination as described in Section 2.5. After applying the ANNs to each pattern in the 4D-DD, a 2D orientation map is obtained and displayed in Fig. 9b. The gradual change in the orientation map indicates continuous bending in the region close to defect within the angular range of ±0.3˚. The angular resolution demonstrated here is consistent with the simulation estimation of 0.0092˚. This is a significant improvement over the angular sensitivity of ~0.3 to 0.8˚ reported using the template matching method based on kinematical simulation [23, 24].

To compare, we applied the pattern matching method to the same dataset. In this method, each experimental diffraction pattern in the 4D-DD is compared with the library of simulated GaSb diffraction patterns to find the best match. First, both of the experimental and simulation diffraction patterns are reduced to a list of integrated intensities following the same procedures we used in the ANN method. Next, the Pearson correlation coefficient between the experimental intensity list and the simulated intensity list is calculated.

$$PCC(x,y) = \frac{\sum_i (x_i - \bar{x})(y_i - \bar{y})}{\sqrt{\sum_i (x_i - \bar{x})^2 \sum_i (y_i - \bar{y})^2}}, \quad (5)$$

where x and y represent experimental and simulated intensities, respectively. While PCC produces similar results as normalized cross-correlation (NCC), it is reported that PCC can be less sensitive to variations in background intensity [32]. This is repeated for each diffraction pattern in the library. The best match is selected based on the largest correlation value. The orientation map is obtained by plotting the best match for each experimental pattern in the 4D-DD (Fig. 9c).

The orientation maps obtained using the two different methods show the same trend but differ in some of the details. One obvious difference is the smoothness of the maps. While the pattern matching results are in the discrete steps of 0.02° with the step size determined by the simulation, the ANN method is able to produce interpolated results, due to the multivariate regression nature of the trained neural network.

### 3.2. Precision orientation mapping of grain subdivision in irradiated $UO_2$

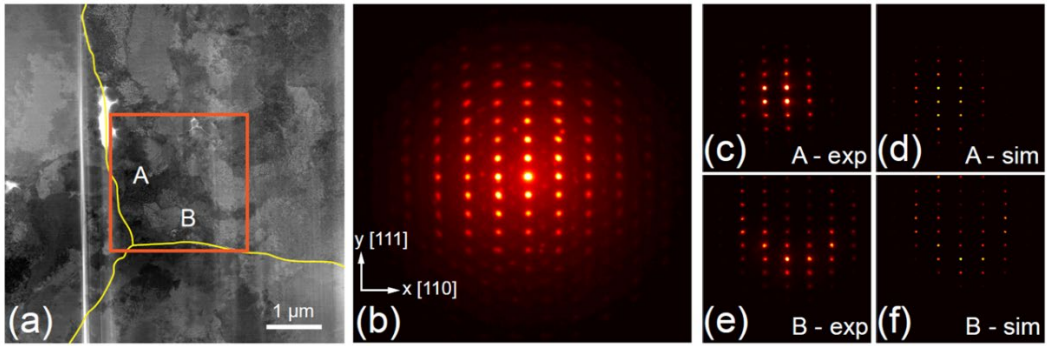

**Fig. 10.** Grain subdivision in irradiated $UO_2$. (a) Annular dark-field image of the sample region investigated by SEND. The yellow lines mark the original grain boundaries in the sample. Orange box denotes the area of 4D-DD acquisition. (b) Virtual selected area diffraction pattern generated from the 4D-DD showing a mosaic diffraction pattern due to grain subdivision. (c)(e) Selected experimental and (d)(f) simulated diffraction patterns with the orientation determined by the trained ANNs.

Grain subdivision is a phenomenon that occurs when the quantity of defects is increased, for example, in cold rolled metals [33] or in nuclear fuel [34, 35]. Subdivision is observed when a grain is divided into many sub grains of smaller size. The size and orientation of sub grains correlate with the amounts of defects and the type of defects. Radiation-induced grain subdivision in $UO_2$ is observed at the rim region of the fuel pellets with high burn-up [35]. The new structure, called high burn-up structure (HBS) or rim structure, is typically composed of sub-micron grains with respect to ~10 μm for the original grains. The main formation mechanism of HBS is still debated and generally considered as either irradiation-induced grain polygonization or grain recrystallization/growth process, featured with the formation of low-angle grain boundaries (LAGBs) and high-angle grain boundaries (HAGBs), respectively [36] Grain orientation or grain boundary nature is often characterized by EBSD and transmission kikuchi diffraction (tKD) in a SEM. Each of these methods offers its unique advantages for materials characterization. TEM offers the highest spatial resolution. Here we demonstrate that the precision can be improved for TEM based diffraction applications.

To demonstrate the potential of ANNs for fine grain orientation mapping, we studied grain subdivision in an irradiated $UO_2$ Sample. $UO_2$ is of interest as a nuclear fuel material. The ADF image in Fig. 10a gives an overview of the sample being investigated. In the thin lamella prepared by FIB, there are three major crystalline grains where the grain boundaries are outlined by yellow lines in Fig. 10a. After irradiation, complex diffraction contrast appears near the grain boundaries indicating changes in the diffraction condition. A 4D-DD is acquired over a region of 2.5×2.5 μm$^2$, with step size of 25 nm, as marked in Fig. 10a. By averaging over all 10,000 diffraction patterns in the 4D-DD, a virtual selected area electron diffraction (SAED) pattern is obtained and shown in Fig. 10b. The SAED shows that the majority of this grain is in [112] zone axis, while the elongation of the diffraction spots indicates a mosaic spread within the crystalline grain, while the uneven contrast in the ADF image indicates the possibility of grain subdivision. By taking a close-up look at two diffraction patterns (Fig. 10c, e) taken from position A and B marked in Fig. 10a, a clear difference in crystal orientation is seen. However, the amount of change indicated by the diffraction patterns is small and around the [112] zone axis. To determine the change in crystal orientation, we applied the trained ANNs for $UO_2$ as described in Section 2.5 to these two patterns. The tilt angles of pattern A (Fig. 10c) from [112] axis are determined to be -0.4155° along X∥[110] direction and -0.4385° along Y∥[111] direction. The tilt angles of pattern B (Fig. 10e) are 0.1081° along X and 1.5119° along Y. The theoretical diffraction patterns of $UO_2$ with these orientations and thickness of 60 nm are simulated under the same condition as the experiment (Fig. 10d, f). The close match between experimental patterns and simulated ones for both A and B shows the accuracy of the ANN method.

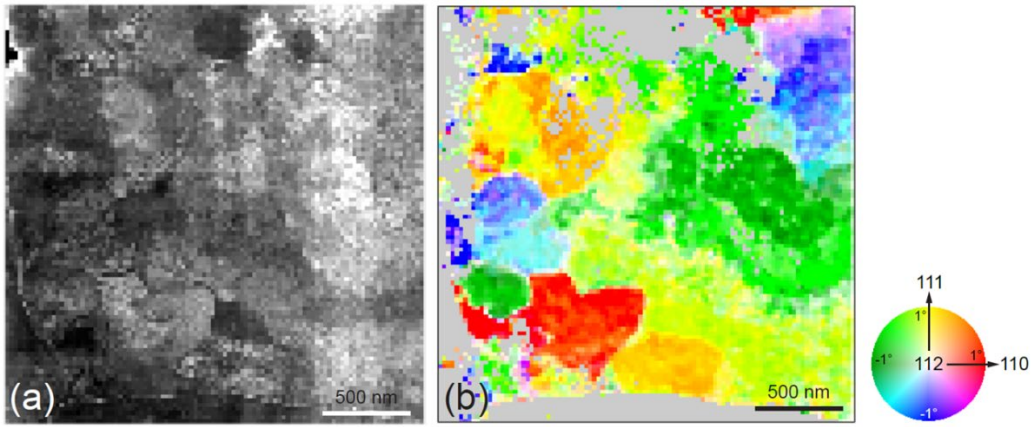

**Fig. 11.** Precision orientation mapping of grain subdivision in irradiated $UO_2$. (a) Virtual annular dark-field image generated from the SEND dataset showing diffraction contrast of multiple sub grains. (b) High-resolution quantitative orientation mapping obtained by ANN method. The gray color in the map indicates regions far away from [112] zone axis.

Then we applied the ANNs to all patterns in the SEND dataset to generate a 2D orientation map as shown in Fig. 11b. Except the gray areas which are from vacuum or different grains with orientation far away from [112] zone, most of sub grains within the scanned area are orientated within 1° from [112]. The map clearly shows the spatial distribution and the size of small grains after subdivision sharing LAGBs with each other. As a comparison, Fig. 11a shows a virtual ADF image generated by integrating all diffracted beams in each pattern of the SEND dataset.

While the diffraction contrast in virtual ADF image provides qualitative information about grain subdivision, our orientation map calculated by the ANN method produces quantitative information with high precision.

### 3.3. Lattice strain mapping of a Si-based FinFET device

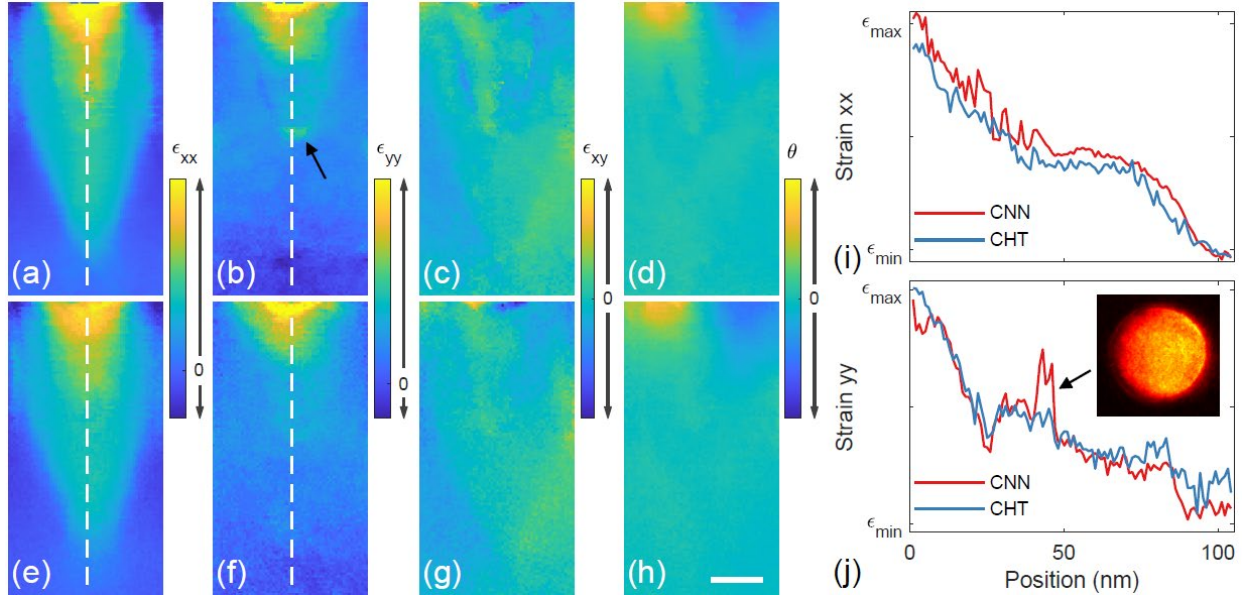

**Fig. 12.** Strain mapping of a Si-based FinFET device. (a-d) Strain maps calculated using CNNs for diffraction disk detection. (e-h) Strain maps calculated using CHT for diffraction disk detection. The scale bar is 20 nm. (i)(j) Strain profiles taken from white dashed lines in (a)(e) and (b)(f), respectively. The color scale indicates the strain values from low (blue) to high (yellow). The inset of (j) displays a distorted diffraction disk taken from the position marked by the black arrow in (b).

To demonstrate the capability of local d-spacing determination using CNNs, the method is applied to a SEND dataset acquired on a sample of Si-based FinFET device of Intel 14 nm technology. The details of the sample and data acquisition are described elsewhere [25]. The SEND map has a size of 60×120 pixel, with pixel (step) size of 1 nm. Each diffraction pattern contains 8 lowest-order diffracted beams in addition to the center beam. 9 sub-images with size of 121×121 pixels containing individual diffraction disks were cropped out from the original diffraction pattern. The trained CNN model described in Section 2.6 was applied to each sub-image to determine the position of the diffraction disk in both horizontal and vertical directions. The measured disk positions were then used to fit a 2D reciprocal lattice. The strain was calculated based on the averaged reference taken from a flat region near the bottom of the scanned area, following the same procedure in our previous paper [25].

All four components of the 2D strain tensor were calculated and displayed in Fig. 12a-d, along with the stain maps calculated using the CHT method for disk detection from the same dataset in Fig. 12e-h. Overall distribution of the strain fields measured by two methods are very close to each other. Line profiles taken from the 2D maps as marked by white dashed lines in Fig. 12a, b, e, and f are displayed in Fig. 12i and j for direct comparison. Strain maps of CNN method appear smoother than those of CHT method, especially in $\epsilon_{yy}$. This improvement is consistent with our estimation of strain measurement precision based on a calibration scan shown in Fig. 8. The most significant deviation between the two results can be seen in Fig. 12j near 45 nm region, marked by black arrows in Fig. 12b and j, where CNN result contains a sudden peak while CHT result is relatively flat. A closer look at the diffraction patterns at that region reveals that the peak in CNN result comes from the presence of additional diffuse diffraction disk (Fig. 12j inset) due to the diffraction at the interface of two SiGe structures of different thickness. When training the CNN model for disk position detection, we only used diffraction patterns simulated with single crystal Si model without considering diffuse scattering. We believe that by training the CNN using simulation from a more complicated and realistic atomic model will help to improve this method.

## 4. Discussions

The above results demonstrated two ANNs based approaches toward data mining of 4D electron diffraction datasets. Such datasets collected by SEND or 4D-STEM techniques provide spatially diffraction information that have revolutionized the characterization of materials microstructure, from the determination of nanodomains in ferroelectrics [37, 38] and molecular frameworks [39] to molecular crystal orientation mapping [22] and to the determination of nanoprecipitates in Al alloys [19]. In all of these applications, efficient and robust data mining techniques play a critical role. Thus, the capability of deep learning and recognizing complex patterns provided by ANNs have the potential to transform how we analyze electron diffraction data, including large 4D-DDs.

How to train ANNs for diffraction data, however, is less explored compared with other areas of electron microscopy. Here we focused on orientation and strain mapping using 4D-DD. In both cases, conventional techniques exist. Strain mapping based on disk position determination using the center of mass or CHT techniques does not require pre-computation, while training ANNs for strain mapping described here is based on simulated diffraction patterns. The advantage of using simulation is that the dynamical effects are taken into consideration. ANNs analysis uses the full intensity and diffraction geometry information, this improves the robustness of the method, and improves the precision in some cases as we demonstrated here.

For precision orientation mapping, both the commonly used pattern matching method and our ANN method require a precalculated database to perform the matching or training, as described in Section 2.5. Here, ANNs hold the advantage in improving the angular resolution by regression through interpolation in high dimensional space capability of ANNs.

A critical factor in training ANNs is to add appropriate noise to the training data. In orientation analysis because of the error that could be introduced by inelastic background for example, we found adding the noise at 15-30 dB level greatly improves the performance of trained ANNs. For strain mapping, the noise is directly added to the image as illustrated in Fig. 2. Here the noise mainly comes from the detector. Adding noise enables the trained ANNs to deal with real experimental data.

The simulation of training datasets and the training of ANNs do take significant amount of time, however, the computational time for applying ANN to the experimental patterns is not related to the size of the simulation library once the ANNs are trained. While for the pattern matching method, computational time is proportional to the simulation library size as each experimental pattern has to be compared with all simulated patterns. Thus, for large 4D-DDs, ANNs can significantly reduce the amount of processing time.

The example of GaSb precision orientation mapping described in Section 3.1 used Matlab with a 3.3GHz CPU. It takes 0.02 s to process 2,500 diffraction patterns in the SEND dataset using the ANN method, while it takes 500 s to finish pattern matching between 2,500 experimental patterns and 13,005 simulated patterns. In this case, the computational time for simulation library was the same for both methods, due to the fact that they shared the same library. To fully take advantage of the high precision provided by the dynamical diffraction simulation, a new library needs to be generated every time when a different material, sample thickness or zone axis is used in an experiment. When the extra sensitivity is not needed, one can also use the kinematically simulated diffraction patterns as the training data for a broader range of sample thicknesses.

For strain mapping, the total processing time is improved by a factor of 4 using CNNs compared with CHT method. The strain precision is also improved as can be seen in Fig. 8 and Fig. 12. One drawback here is the amount of computational time for pre-computation, but once ANNs are trained, the same network can be applied repeatedly to datasets acquired with similar experimental conditions. We note that by training the ANNs with the dynamically simulated diffraction disks, they can automatically learn to focus on circular edges similar to the human designed CHT method. Therefore, if a large enough training dataset is built to take into account variations in the diffraction disks, it is possible to train a single set of CNNs capable of measuring strain from different datasets of different samples.

In choosing the right neural networks for our applications, we have kept the network design as simple as possible. This approach is based on the consideration that diffraction data are highly structured and thus the diversity of images is significantly lower than for example, random taken images from a heterogeneous TEM sample. We have used ANNs and CNNs here for our applications, Once the type of NNs is selected, the detailed choices for the NNs are the number of layers, number of neurons in each layer. In the convolutional NNs, the convolution is performed using eight 3×3 filters repeated two times and with 2×2 max pooling in between. The filter size can be increased here as well. Compared to image recognition where a large number of neurons are required, the number of neurons for diffraction analysis is modest. For example, in the CNN used for strain analysis, the total number of neurons is 800k and the training time on a desktop computer without GPU is only about 30 mins, which provides a real advantage for applications. For ANNs, one-hidden-layer structure with 30 neurons has 1,300 parameters. The training takes about 10 minutes using a desktop PC. Thus, the modest computational requirement for diffraction pattern analysis is another benefit for electron diffraction.

For future developments, we note the rapid developments in electron detector technology with faster frame acquisition, large dynamical range and improvement in detective quantum efficiency (DQE) using direct electron detection [40]. All of these improve the diffraction quality in term of signal noise ratio, as well as the type of diffraction patterns that can be collected, and crystals that can be studied. Thus, we expect the improvement in measurement precision and computational time using machine learning will play key enabling role for 4D-DD analysis.

## 5. Conclusions

We have provided two examples of machine learning assisted electron diffraction pattern analysis, and how such analysis is incorporated with scanning electron nanodiffraction. The first example is a simple artificial neural network designed to determine crystal orientation based on the integrated diffraction disk intensities instead of the whole pattern. We demonstrated that it is possible to achieve faster and more accurate determination of small orientation change in a GaSb thin sample compared with the traditional correlation-based pattern matching. The method is applied to characterize the misorientation of grain subdivision in a sample of irradiated $UO_2$. The results clearly show the spatial distribution of multiple small grains sharing low-angle grain boundaries less than 2° with each other. The second example is a convolutional neural network designed to measure diffraction disk position from the pattern. Since this method works on small sub-images, the network structure can be simplified to expedite both model training and processing of the experimental data. The application of the trained convolutional neural network to the measurement of strain fields in a FinFET device shows comparable results as previously calculated by the circular Hough transform method and has better precision in some cases. The training of all these neural networks is possible with accurate electron diffraction simulation using dynamical diffraction theory. Together, supervised machine learning based automated analysis of large 4D diffraction datasets can provide rich information about nanoscale crystalline materials with both high spatial resolution and high precision.

**Acknowledgments**
This work is supported by Intel Corporation through an SRC project (Award #54071821) and Idaho National Laboratory through a LDRD project (LDRD # 19A42-017FP). The source code used in this paper is available at https://github.com/flysteven/ANN-4DDD.